\batchmode

\documentstyle[12pt] {article}

\def\zid{1\kern-0.36em\llap~1}

\newcommand{\beq}{\begin{equation}}
\newcommand{\ber}{\begin{eqnarray}}
\newcommand{\eeq}{\end{equation}}
\newcommand{\eer}{\end{eqnarray}}

\begin{document}

\begin{titlepage}
\rightline{[SUNY BING 7/1/95] }
\vspace{2mm}
\begin{center}
{\large \bf A GENERAL TREATMENT OF TAU SEMI-
LEPTONIC
DECAYS}\\[2mm]
Charles A. Nelson\footnote{Electronic address: cnelson @
bingvmb.cc.binghamton.edu .  Contributed paper to LP95, Beijing.
\newline Expanded version to be submitted to Phys. Rev.}, Minseob
Kim, Daryl P. Nazareth and Hui-Chun
Yang \\
{\it Department of Physics\\State University of New York at
Binghamton\\
Binghamton, N.Y. 13902-6016}\\[5mm]
\end{center}


\begin{abstract}

The most general Lorentz invariant spin-correlation functions for
$\tau
^{-}\rightarrow \rho ^{-}\nu ,a_1^{-}\nu ,K^{*-}\nu , \newline \pi
^{-}\nu ,K^{-}\nu $%
\ are expressed in terms of eight semi-leptonic parameters. The
parameters
are physically defined in terms of tau-decay partial-width-intensities
for
polarized-final-states.  The parameters
are also expressed in terms of a ``$(V-A)\ +\ $additional chiral
coupling''
structure in the ${J^{Charged}}_{Lepton}$ current, so as to
bound the scales $\Lambda $\ for ``new physics'' such as arising
from
leptonic CP violation, tau weak magnetism, weak electricity, and/or
second-class currents. The two tests for leptonic CP violation in
$\tau
\rightarrow \rho \nu \ $ decay are generalized to $\tau \rightarrow
a_1\nu \
$ decay and to two additional tests if there are $\nu _R\ $and $\bar
\nu _L\
$couplings.
\end{abstract}

\newpage

\section{INTRODUCTION}

The principal purpose of this paper is to provide a general treatment
of
two-body tau decays \cite{0}which only assumes Lorentz
invariance
and exploits the tree-like structure of the dominant contributions to
the $\tau ^{-}\tau ^{+}$
production-decay sequence. In particular, CP invariance and a
$(V\mp A)\ $%
structure of the tau charged-current is not assumed. In a separate
paper  \cite{1}, it
has been reported that by means of the associated stage-two spin-
correlation
functions$\ $the scales of $\Lambda \approx \ few$\ $100GeV\
$can
be probed
at $M_Z$ center-of-mass energy in unpolarized $e^{-}e^{+}$\
collisions. The
scale of $1-2TeV\ $ can be probed at $10GeV$\ or $4GeV$.

Previously in the study of the weak-interaction's charged-current in
muonic
and in hadronic processes, it has been important to determine the
complete
Lorentz structure directly from experiment in a model independent
manner.
Here, in Sec. 2, eight semi-leptonic parameters are defined for a
specific
tau semi-leptonic decay mode such as $\tau ^{-}\rightarrow \rho ^{-
}\nu \ $.
The parameters are physically defined in terms of tau-decay
partial-width-intensities for polarized-final-states. They can also be
simply expressed in terms of the helicity amplitudes $A(\lambda
_\rho
,\lambda _\nu )\ $for $\tau ^{-}\rightarrow \rho ^{-}\nu \ $.

Besides model independence, a major current issue is whether or
not
there is an additional chiral coupling in the tau's charged-current. A
chiral classification of
additional structure is a natural phenomenological extension of the
symmetries of the standard $SU(2)_L\ X\ U(1)$\ electroweak lepton
model. The
requirement of $\bar u(p_\nu )\rightarrow \bar u(p_\nu )\frac
12(1+\gamma _5)
$ and/or $u(k_\tau )\rightarrow \frac 12(1-\gamma _5)u(k_\tau )$\
invariance
of the vector and axial current matrix elements $\langle \nu \left|
v^\mu
(0)\right| \tau \rangle $\ and $\langle \nu \left| a^\mu (0)\right| \tau
\rangle $,$\ $allows only $g_L,g_{S+P},g_{S^{-}+P^{-
},}g_{+}=f_M+f_E,$and $%
\tilde g_{+}=T^{+}+T_5^{+}\ $couplings. From this $SU(2)_L$
perspective, the
relevant experimental question is what are the best current limits on
such
additional couplings? Similarly, $\bar u(p_\nu )\rightarrow \bar
u(p_\nu
)\frac 12(1-\gamma _5)$ and/or $u(k_\tau )\rightarrow \frac
12(1+\gamma _5)\
u(k_\tau )$\ invariance selects the complimentary set of $%
g_R,g_{S-P},g_{S^{-}-P^{-},}g_{-}=f_M-f_E,$and $\tilde g_{-
}=T^{+}-T_5^{+}\ $%
couplings. The absence of $SU(2)_R$ couplings is simply built into
the
standard model; it is not predicted by it. So, what are the best
current
limits on such $SU(2)_R$ couplings in tau physics?

In Sec.3, as a step towards precision answers to these basic
questions, the
semi-leptonic parameters are expressed in terms of a ``$(V-A)+$\
additional
chiral coupling'' structure in the ${J^{Charged}}_{Lepton}$
current \cite{1}. Two tables display the resulting values of the
parameters when the
various additional chiral couplings $(g_i/2\Lambda _i)\ $are small
relative
to the standard $V-A$\ coupling $(g_L).$\

Sec. 4 gives the most general Lorentz invariant spin-correlation
functions
for $e^{-}e^{+}\rightarrow \tau ^{-}\tau ^{+}$\ followed by $\tau
\rightarrow \rho \nu ,a_1\nu ,K^{*}\nu $\ including both $\nu
_{L,R}$\
helicities and both $\bar \nu _{R,L}$\ helicities.\ These same
parameters
appear in the general angular distributions
\beq
\frac{dN}{d(\cos \theta_1^\tau )d(\cos \tilde \theta _a)d \tilde
\phi_a} = {\bf R}_{\pm \pm }
\eeq
for the polarized $\tau^{-}$ decay
chain, see Ref. \cite{C94}.  So, they can also
be directly measured by means of longitudinally-polarized beams at
a
tau/charm factory or at a B-factory with longitudinally polarized
beams.

In Sec. 5, the two tests for leptonic CP violation in $\tau \rightarrow
\rho
\nu \ $ decay are generalized to $\tau \rightarrow a_1\nu \ $ decay
and to
two additional tests if there are $\nu _R\ $and $\bar \nu _L\
$couplings \cite{C94a}. Sec.6 treats $\tau ^{-}\rightarrow \pi ^{-
}\nu ,K^{-}\nu $\ decay. These modes each provide less
information
since here only two of the semi-leptonic parameters can be
measured. The fundamental $S^{-}$\ and $P^{-}$\ couplings do
not
contribute to $\tau \rightarrow \rho \nu ,a_1\nu ,K^{*}\nu $\ but
they are also found to be suppressed in $\tau ^{-}\rightarrow \pi ^{-
}\nu
,K^{-}\nu $\ decay$.$ In the appendix we list the $A(\lambda _\rho
,\lambda
_\nu )\ $for $\tau ^{-}\rightarrow \rho ^{-}\nu $\ for the most
general tau ${J^{Charged}}_{Lepton}$ current.

\section{PARAMETRIZATION OF TAU SEMI-LEPTONIC \protect\newline DECAY MODES}

The reader should be aware that it is not necessary to use the
helicity
formalism \cite{5} because the parameters introduced below will be
fundamentally defined in terms tau-decay partial width intensities
for
polarized-final-states. However, the helicity
formalism does provide a lucid, neat, and
flexible framework for connecting the most general Lorentz
invariant
couplings at the Lagrangian level, for describing tau lepton
decays, with the most general Lorentz invariant spin-correlation
functions for $\tau^- \tau^+$ pair production.   In practice, the
helicity formalism also frequently provides insights and checks on
the
resulting formulas and their symmetries.  We present the discussion
for the $\rho
\nu$ channel, but the
same formulas hold for the $a_1 \nu$ and $K^* \nu$ channels.

In the  $\tau ^{-}$ rest frame, the
matrix element for $\tau ^{-}\rightarrow \rho ^{-}\nu$ is
\beq
\langle \theta _1^\tau ,\phi _1^\tau ,\lambda _\rho ,\lambda _\nu
|\frac
12,\lambda _1\rangle =D_{\lambda _1,\mu }^{\frac 12*}(\phi
_1^\tau ,\theta
_1^\tau ,0)A\left( \lambda _\rho ,\lambda _\nu \right)
\eeq
where $\mu =\lambda _\rho -\lambda _\nu $ and $\lambda_1$ is the
$\tau^{-}$
helicity.   For the $CP$-conjugate
process, $\tau
^{+}\rightarrow \rho ^{+}\bar \nu \rightarrow \left( \pi ^{+}\pi
^o\right) \bar \nu $, in the $\tau ^{+}$ rest frame
\beq
\langle \theta _2^\tau ,\phi _2^\tau ,\lambda _{\bar \rho },\lambda
_{\bar
\nu }|\frac 12,\lambda _2\rangle =D_{\lambda _2,\bar \mu }^{\frac
12*}(\phi
_2^\tau ,\theta _2^\tau ,0)B\left( \lambda _{\bar \rho },\lambda
_{\bar \nu
}\right)
\eeq
with $\bar \mu =\lambda _{\bar \rho }-\lambda _{\bar \nu }$.

These formulas only assume Lorentz invariance and do not assume
any
discrete symmetry properties.  Therefore, it is easy to use this
framework for
testing for the consequences of such addtional symmetries.  In
particular,  for  $%
\tau ^{-}\rightarrow \rho ^{-}\nu $ and $\tau ^{+}\rightarrow \rho
^{+}\bar
\nu $ a specific discrete symmetry implies a specific relation among
the associated
helicity amplitudes:%
$$
\begin{array}{cc}
\underline{Invariance} & \underline{Relation} \\ P & A\left( -
\lambda _\rho
,-\lambda _\nu \right) =A\left( \lambda _\rho ,\lambda _\nu \right)
\\
& B\left( -\lambda _{\bar \rho },-\lambda _{\bar \nu }\right)
=B\left(
\lambda _{\bar \rho },\lambda _{\bar \nu }\right)  \\
C & B\left( \lambda _{\bar \rho },\lambda _{\bar \nu }\right)
=A\left(
\lambda _{\bar \rho },\lambda _{\bar \nu }\right)  \\
CP & B\left( \lambda _{\bar \rho },\lambda _{\bar \nu }\right)
=A\left(
-\lambda _{\bar \rho },-\lambda _{\bar \nu }\right)  \\
\tilde T_{FS} & A^{*}\left( \lambda _\rho ,\lambda _\nu \right)
=A\left(
\lambda _\rho ,\lambda _\nu \right)  \\
& B^{*}\left( \lambda _{\bar \rho },\lambda _{\bar \nu }\right)
=B\left(
\lambda _{\bar \rho },\lambda _{\bar \nu }\right)  \\
CP\tilde T_{FS} & B^{*}\left( \lambda _{\bar \rho },\lambda
_{\bar
\nu
}\right) =A\left( -\lambda _{\bar \rho },-\lambda _{\bar \nu }\right)
\end{array}
$$
Measurement of a non-real helicity
amplitude implies
a violation of $\tilde T_{FS}$ invariance when a first-order
perturbation in an
``effective" hermitian Hamiltonian is reliable.  So $\tilde T_{FS}$
invariance is
expected to be violated when there are significant final-state
interactions; and it is
to be distinguished from canonical $T$ invariance which requires
interchanging
``final'' and ``initial'' states, i.e. actual time-reversed reactions are
required.

\subsection{Definition by partial width intensities for
polarized-final-states}

The tau semi-leptonic decay parameters for $\tau ^{-}\rightarrow
\rho
^{-}\nu $, and likewise for $\tau ^{-}\rightarrow {a_1}^{-}\nu $
and
\newline $\tau
^{-}\rightarrow {K^*}\nu $ , are defined by
\begin{equation}
\begin{array}{c}
\zeta \equiv (\Gamma _L^{-}-\Gamma _T^{-})/(
{\cal S}_\rho \Gamma ) \\ \sigma \equiv (\Gamma _L^{+}-\Gamma
_T^{+})/(
{\cal S}_\rho \Gamma ) \\ \xi \equiv \frac 1\Gamma (\Gamma _L^{-
}+\Gamma
_T^{-})
\end{array}
\end{equation}
where $\Gamma \equiv \Gamma _L^{+}+\Gamma _T^{+}$ is the
total partial width
for $\tau ^{-}\rightarrow \rho ^{-}\nu $. The subscripts denote the
polarization of the final $\rho ^{-}$, either ``L=longitudinal'' or
``T=transverse'', and the superscripts denote ``$\pm $ for
sum/difference of
the $\nu _{L\ }$versus $\nu _R$ contributions''. Such final-state-
polarized
partial widths are in principle physical observables but their direct
measurement would require measurement of the polarizations of
both the final
$\rho ^{-}$ and $\nu $. In Sec. 4 below, we will explain how the
equivalent semileptonic parameters can be measured by various
spin-
correlation
techniques.

To be clear about the terminology and sign-conventions, note that in
terms of the helicity amplitudes $A(\lambda _\rho ,\lambda _\nu )$
these
final-state-polarized partial widths are:
\begin{equation}
\begin{array}{c}
\Gamma _L^{\pm }=\left| A(0,-\frac 12)\right| ^2\pm \left| A(0,\frac
12)\right| ^2 \\
\Gamma _T^{\pm }=\left| A(-1,-\frac 12)\right| ^2\pm \left| A(-
1,\frac
12)\right| ^2
\end{array}
\end{equation}
Recall \cite{C94} that by rotational invariance the other $\rho ^{-}$
helicity amplitudes are
forbidden; similarly for the $\rho ^{+}\
$mode in $%
\tau ^{+}\ $decay, the $B(1, \frac 12)$ and $B(-1,-\frac 12)$
amplitudes vanish.

To describe the contributions from the interference between the
longitudinal(%
$L$) and transverse($T$) vector-meson amplitudes in the decay
process, the
additional parameters are:
\begin{equation}
\begin{array}{c}
\omega \equiv I_{
{\cal R}}^{-}\ /({\cal R}_\rho \Gamma ) \\ \eta \equiv I_{
{\cal R}}^{+}\ /({\cal R}_\rho \Gamma ) \\ \omega ^{\prime
}\equiv
I_{
{\cal I}}^{-}\ /({\cal R}_\rho \Gamma ) \\ \eta ^{\prime }\equiv
I_{{\cal I}%
}^{+}\ /({\cal R}_\rho \Gamma )
\end{array}
\end{equation}
In terms of the helicity amplitudes these measurable $LT$-
interference
intensities are
\begin{equation}
\begin{array}{c}
I_{
{\cal R}}^{\pm }={\cal RE}\{A(0,-\frac 12)^{*}A(-1,-\frac 12)\pm
A(0,\frac
12)^{*}A(1,\frac 12)\} \\ =\left| A(0,-\frac 12)\right| \left| A(-1,-\frac
12)\right| \cos \beta _a \\
\pm \left| A(0,\frac 12)\right| \left| A(1,\frac 12)\right| \cos \beta
_a^R
\end{array}
\end{equation}
\begin{equation}
\begin{array}{c}
I_{
{\cal I}}^{\pm }={\cal IM}\{A(0,-\frac 12)^{*}A(-1,-\frac 12)\pm
A(0,\frac
12)^{*}A(1,\frac 12)\} \\ =\left| A(0,-\frac 12)\right| \left| A(-1,-\frac
12)\right| \sin \beta _a \\
\pm \left| A(0,\frac 12)\right| \left| A(1,\frac 12)\right| \sin \beta
_a^R
\end{array}
\end{equation}
where $\beta _a\equiv \phi _{-1}^a-\phi _0^a$, $\beta _a^R\equiv
\phi
_1^a-\phi _0^{aR}$\ are the measurable phase differences of of the
associated helicity amplitudes $A=\left| A\right| \exp \iota \phi $.\

Note that the hadronic factors ${\cal S}_\rho $ and ${\cal R}_\rho $
do
depend on the particular tau semi-leptonic decay channel. For the
$\rho $
mode they are given by
\begin{equation}
{\cal S}_\rho =\frac{1-2\frac{m_\rho ^2}{m^2}}{1+2\frac{m_\rho
^2}{m^2}}
\end{equation}
\begin{equation}
{\cal R}_\rho =\frac{\sqrt{2}\frac{m_\rho }m}{1+2\frac{m_\rho
^2}{m^2}}
\end{equation}
In this section, these ${\cal S}_\rho $ and ${\cal R}_\rho $ factors
have
been explicitly inserted into the definitions of the semi-leptonic
decay
parameters, so that quantities such as ${q_\rho }^2={m_\rho }^2$
can be
smeared over in application of the spin-correlation functions given
below.  Such smearing is needed due to the finite $\rho $ width.
This
treatment assumes that the momentum
dependence (i.e. the dependence on ${q_\rho }^2$,$\ldots $ ) of the
form-factors $g_L$ and $g_i$ is negligible. Depending on the
application and
on the desired experimental test, more sophisticated treatments of
the
${%
q_\rho }^2$,$\ldots $ dependence should be used such as ones
which
incorporate results from recent QCD calculations for tau decays
\cite{6} and ones which include possible contributions
from addtional resonances such as the $\rho ^{\prime }$. Because
of
the
smearing and the good understanding of QCD in tau physics, we do
not expect
this to be a fundamental difficulty in practice but rather a technical
matter that
requires sufficient care.

For the $a_1,K^{*}\ $modes, replace respectively $m_\rho
\rightarrow
m_{a_1},m_{K^{*}}$. These factors numerically are $({\cal
S},{\cal R})_{\rho
,a_1,K^{*}}=0.454,0.445;-0.015,0.500;0.330,0.472.\ \ $Recall
\cite{7} that ${\cal S}_{\pi ,K}=1$\ for $J=0$, so ${\cal S}_{\rho
,a_1,K^{*}}$%
suppresses the spin signatures when $J\neq 0$. On the other hand,
${\cal R}%
_{\rho ,a_1,K^{*}}\ $doesn't appear for $J=0\ $channels since their
sequential-decay-chains end with the first stage.

Depending on the physics and/or experimental situation, it may
sometimes be
advantagous to rewrite the spin-correlation function(s) of interest
directly
in terms of the above final-state-polarized partial widths and
$LT$%
-interference intensities, instead of using the above $\tau $ semi-
leptonic
decay parameters.

For the CP conjugate modes, $\tau ^{+}\rightarrow \rho ^{+}\bar
\nu
$ and $%
\tau ^{+}\rightarrow {a_1}^{+}\bar \nu $, the formulas for their
semi-leptonic decay parameters are the same except that all
quantities are
``barred,'' and there is the substitution of helicity amplitudes $%
A(x,y)\rightarrow B(-x,-y)$. This parametrization only assumes
Lorentz
invariance; for example, a simple test of CP invariance is that each
''barred'' semi-leptonic parameter is measured to be equal to it's
``unbarred'' associate (within experimental
errors).

\section{Significance of semi-leptonic parameters versus \newline
``Chiral Couplings"}

The most general Lorentz coupling for \hskip
1em  $\tau^{-
}\rightarrow \rho
^{-}\nu _{L,R}$ is
\beq
\rho _\mu ^{*}\bar u_{\nu _\tau }\left( p\right) \Gamma ^\mu
u_\tau
\left(
k\right)
\eeq
where $k_\tau =q_\rho +p_\nu $. It is convenient to treat the vector
and
axial vector matrix elements separately. In Eq.(11)
\beq
\Gamma _V^\mu =g_V\gamma ^\mu +
\frac{f_M}{2\Lambda }\iota \sigma ^{\mu \nu }(k-p)_\nu   +
\frac{g_{S^{-}}}{2\Lambda }(k-p)^\mu +\frac{g_S}{2\Lambda
}(k+p)^\mu
+%
\frac{g_{T^{+}}}{2\Lambda }\iota \sigma ^{\mu \nu }(k+p)_\nu
\eeq
\beq
\Gamma _A^\mu =g_A\gamma ^\mu \gamma _5+
\frac{f_E}{2\Lambda }\iota \sigma ^{\mu \nu }(k-p)_\nu \gamma
_5
+
\frac{g_{P^{-}}}{2\Lambda }(k-p)^\mu \gamma
_5+\frac{g_P}{2\Lambda }%
(k+p)^\mu \gamma _5  +\frac{g_{T_5^{+}}}{2\Lambda }\iota
\sigma ^{\mu \nu
}(k+p)_\nu \gamma _5
\eeq

The parameter
$%
\Lambda =$ ``the scale of New Physics''. In effective field theory
this
is the scale at which new particle thresholds are expected to
occur or where the theory becomes non-perturbatively strongly-
interacting so as to overcome perturbative inconsistencies.  In old-
fashioned renormalization theory $\Lambda$  is the scale at
which the calculational methods and/or the principles of
``renormalization''
breakdown, see for example \cite{th}. While some terms of the
above do occur as higher-order perturbative-corrections in the
standard model,
such SM
contributions  are ``small'' versus the sensitivities of present tests in
$\tau$ physics
in the analogous cases of the $\tau$'s neutral-current and
electromagnetic-current
couplings, c.f.
\cite{d0}.  For charged-current couplings, the situation should be
the
same.

Without additional theoretical, c.f. \cite{1}, or experimental
inputs, it is not possible to select what is the "best" minimal set of
couplings for
analyzing the structure of the tau's charged current.  For instance,
by
Lorentz
invariance, there are the equivalence theorems that for the
vector
current%
\ber
S\approx V+f_M, & T^{+}\approx -V+S^{-}
\eer
\noindent
and for the axial-vector current
\ber
P\approx -A+f_E, & T_5^{+}\approx A+P^{-}
\eer
The matrix elements of the divergences of these charged-currents
are
\beq
(k-p)_\mu V^\mu =[g_V(m_\tau -m_\nu )
+
\frac{g_{S^{-}}}{2\Lambda }q^2+\frac{g_S}{2\Lambda }(m_\tau
^2-m_\nu ^2)
 +%
\frac{g_{T^{+}}}{2\Lambda }(q^2-[m_\tau -m_\nu ]^2)]\bar u_\nu
u_\tau
\eeq
\beq
(k-p)_\mu A^\mu =[- g_A(m_\nu +m_\tau )
+
\frac{g_{P^{-}}}{2\Lambda }q^2+\frac{g_P}{2\Lambda }(m_\tau
^2-m_\nu ^2)
 +%
\frac{g_{T_5^{+}}}{2\Lambda }(q^2-[m_\tau+m_\nu ]^2)]\bar
u_\nu \gamma
_5u_\tau
\eeq
Both the weak magnetism  $\frac{f_M}{2\Lambda }$ and the weak
electricty $%
\frac{f_E}{2\Lambda }$ terms are divergenceless. On the other
hand, since $%
q^2=m_\rho ^2$,  even when $m_\nu =m_\tau $ there are non-
vanishing
terms due to
the couplings $S^{-},T^{+},A,P^{-},T_5^{+}$.

\subsection{Semi-leptonic parameters' form in terms of $g_L$ plus
an \newline ``additional chiral coupling''}

We first display the expected forms for the above semi-leptonic
parameters
for the $\tau \rightarrow \rho \nu ,\ a_1\nu ,K^{*}\nu $ \ decay
modes for
the case of a pure $V-A$ chiral coupling as in the standard lepton
model. We assume that the mass of the tau neutrino and anti-
neutrino are negligible, see Table 1 in \cite{e4}. Next we will give
the form for the case of a single chiral coupling $%
(g_i/2\Lambda _i)$\ in addition to the standard $V-A$ coupling. In
this
case, we first list the formula for an arbitrarily large additional
contribution. In two separate tables we list the formulas assuming
that the
additional contribution is small versus the $V-A$ coupling.
Throughout this
paper, we usually suppress the entry in the ``$i$'' subscript on the
new-physics coupling-scale ``$\Lambda _i$'' when it is obvious
from
the
context of interest.

In the case of ``multi-additional'' chiral contributions, the general
formulas for $A(\lambda _\rho ,\lambda _\nu )$\ \ which are listed
in
the
appendix can be substituted into the above definitions so as to
derive
the
expression(s) for the ``multi-additional'' chiral contributions.
Frequently we will suppress the subscript on $m_{\tau}$.

{\em Pure }$V-A$ {\em coupling:}
\begin{equation}
\begin{array}{cc}
\zeta =\sigma =\omega =\eta =\xi =1 \\ \omega ^{\prime }= \eta
^{\prime }=0
\end{array}
\end{equation}

$V+A{\em \ also\ present:}$%
\begin{equation}
\begin{array}{cc}
\zeta = \xi  & \omega =\xi  \\ \sigma =1  & \eta =
1  \\ \xi =\frac{\left| g_L\right| ^2-\left| g_R\right| ^2}{%
\left| g_L\right| ^2+\left| g_R\right| ^2} & \omega ^{\prime }=\eta
^{\prime
}=0
\end{array}
\end{equation}

$S+P\ {\em also\ present:}$%
\begin{equation}
\zeta =\sigma =\left(
\begin{array}{c}
(1-2
\frac{m_\rho ^2}{m^2})\left| g_L\right| ^2+\frac m\Lambda [1-
\frac{m_\rho ^2%
}{m^2}]{\cal RE}(g_L^{*}g_{S+P}) \\ +\{\frac m{2\Lambda }[1-
\frac{m_\rho ^2}{%
m^2}]\}^2\left| g_{S+P}\right| ^2
\end{array}
\right) / ( {\cal S}_\rho {\cal D}^{+} )
\end{equation}
\begin{equation}
\xi =1
\end{equation}
\begin{equation}
\begin{array}{c}
\omega =\eta =
\sqrt{2}\frac{m_\rho }m\left( \left| g_L\right| ^2+\frac m{2\Lambda
}[1-%
\frac{m_\rho ^2}{m^2}]{\cal RE}(g_L^{*}g_{S+P})\right) / ( {\cal
R}_\rho {\cal D}^{+} ) \\ \omega
^{\prime }=\eta ^{\prime }=-\sqrt{2}\frac{m_\rho }{2\Lambda }[1-
\frac{m_\rho
^2}{m^2}]{\cal IM}(g_L^{*}g_{S+P})/ ( {\cal R}_\rho {\cal
D}^{+} )
\end{array}
\end{equation}
where
$$
{\cal D}^{+ }=(1+2\frac{m_\rho ^2}{m^2})\left| g_L\right|
^2+\frac
m\Lambda [1-%
\frac{m_\rho ^2}{m^2}]{\cal RE}(g_L^{*}g_{S+P})+\{\frac
m{2\Lambda }[1-\frac{%
m_\rho ^2}{m^2}]\}^2\left| g_{S+P}\right| ^2
$$

$S-P\ {\em also\ present:}$%
\begin{equation}
\zeta ,\sigma =\left( (1-2\frac{m_\rho ^2}{m^2})\left| g_L\right|
^2\mp
\{\frac m{2\Lambda }[1-(\frac{m_\rho ^2}{m^2})]\}^2\left| g_{S-
P}\right|
^2\right) / ( {\cal S}_\rho {\cal D}^{-} )
\end{equation}
where the upper(lower) sign on the ``rhs'' goes with the
first(second)
entry
on the ``lhs.''
\begin{equation}
\xi =1
\end{equation}
\begin{equation}
\omega =\eta =\sqrt{2}\frac{m_\rho }m\left| g_L\right| ^2/( {\cal
R}_\rho {\cal D}^{-} ),\ \
\omega ^{\prime }=\eta ^{\prime }=0
\end{equation}
where
$$
{\cal D}^{-}=(1+2\frac{m_\rho ^2}{m^2})\left| g_L\right|
^2+\{\frac
m{2\Lambda }[1-\frac{m_\rho ^2}{m^2}]\}^2\left| g_{S-P}\right|
^2
$$

$f_M+f_E\ {\em also\ present:}$

For this case we write the coupling constant of the sum of the weak
magnetism and the weak electricity couplings as
$$
g_{+}=f_M+f_E
$$
In this notation,
\begin{equation}
\zeta =\sigma =\left(
\begin{array}{c}
(1-2
\frac{m_\rho ^2}{m^2})\left| g_L\right| ^2+\frac{m_\rho
^2}{m\Lambda }{\cal %
RE}(g_L^{*}g_{+}) \\ +\frac{m_\rho ^2}{4\Lambda ^2}[-
2+\frac{m_\rho ^2}{m^2}%
]\left| g_{+}\right| ^2
\end{array}
\right) / ( {\cal S}_\rho {\cal D_T}^{+} )
\end{equation}
$$
\xi =1
$$
\begin{equation}
\begin{array}{c}
\omega =\eta =
\sqrt{2}\frac{m_\rho }m\left( \left| g_L\right| ^2-\frac m{2\Lambda
}[1+%
\frac{m_\rho ^2}{m^2}]{\cal RE}(g_L^{*}g_{+})+\frac{m_\rho
^2}{4\Lambda ^2}%
\left| g_{+}\right| ^2\right) / ( {\cal R}_\rho {\cal D_T}^{+} )  \\
\omega ^{\prime }=\eta
^{\prime }=-\frac{m_\rho }{\sqrt{2}\Lambda }[1-\frac{m_\rho
^2}{m^2}]{\cal IM%
}(g_L^{*}g_{+})/ ( {\cal R}_\rho {\cal D_T}^{+} )
\end{array}
\end{equation}
where
$$
{\cal D_T}^{+}=(1+2\frac{m_\rho ^2}{m^2})\left| g_L\right| ^2-
3\frac{m_\rho ^2%
}{m\Lambda }{\cal RE}(g_L^{*}g_{+})+\frac{m_\rho
^2}{4\Lambda ^2}[2+\frac{%
m_\rho ^2}{m^2}]\left| g_{+}\right| ^2
$$
$f_M-f_E\ {\em also\ present:}$

Similarly, we write the coupling constant of the difference of the
weak
magnetism and the weak electricity couplings as
$$
g_{-}=f_M-f_E
$$
and so,
\begin{equation}
\zeta ,\sigma =\left( (1-2\frac{m_\rho ^2}{m^2})\left| g_L\right|
^2\pm
\frac{m_\rho ^2}{4\Lambda ^2}\left| g_{-}\right| ^2\right) / ( {\cal
S}_\rho {\cal D}_T^{-} )
\end{equation}
where the upper(lower) sign on the ``rhs'' goes with the
first(second)
entry
on the ``lhs.''Also,
\begin{equation}
\xi =\left( (1+2\frac{m_\rho ^2}{m^2})\left| g_L\right| ^2-
3\frac{m_\rho ^2}{%
4\Lambda ^2}\left| g_{-}\right| ^2\right) /  {\cal D}_T^{-}
\end{equation}
\begin{equation}
\omega ,\eta =\sqrt{2}\frac{m_\rho }m\left( \left| g_L\right| ^2\mp
\frac{%
m_\rho ^2}{4\Lambda ^2}\left| g_{-}\right| ^2\right) / ( {\cal
R}_\rho {\cal D}_T^{-} ) ,\ \
\omega ^{\prime }=\eta ^{\prime }=0\
\end{equation}
Here%
$$
{\cal D}_T^{-}=(1+2\frac{m_\rho ^2}{m^2})\left| g_L\right|
^2+3\frac{m_\rho
^2}{4\Lambda ^2}\left| g_{-}\right| ^2
$$

$T^{+}+T_5^{+}\ {\em also\ present:}$

We let
$$
\tilde g_{+}=g_{T+T_5}^{+}
$$
In this notation,
\begin{equation}
\zeta =\sigma =\xi=1
\end{equation}
Also
\begin{equation}
\omega =\eta =1 ;\ \ \omega ^{\prime }=\eta ^{\prime
}=0
\end{equation}
A single additional \ $\tilde g_{+}={g^{+}}_{T^{+}+T_5^{+}}\
$coupling does not
change the values from that of the pure $V-A$\ coupling.

$T^{+}-T_5^{+}\ {\em also\ present:}$

\begin{quotation}
We let
$$
\tilde g_{-}=g_{T-T_5}^{+}
$$
and so,
\begin{equation}
\zeta = \xi ,\ \ \sigma =1
\end{equation}
\begin{equation}
\xi =\frac{\left| g_L\right| ^2-\left| \frac{m\tilde g_{-}}{2\Lambda
}%
\right| ^2}{\left| g_L\right| ^2+\left| \frac{m\tilde g_{-}}{2\Lambda
}%
\right| ^2}
\end{equation}
\begin{equation}
\omega = \xi ,\ \ \eta =1,\ \omega ^{\prime }=\eta
^{\prime }=0\
\end{equation}
\end{quotation}
A single additional \ $\tilde g_{-}={g^{+}}_{T^{+}-T_5^{+}}\
$coupling is
equivalent to a single additional $V+A$\ coupling, except for the
interpretation of their respective chirality parameters.

\subsection{ Semi-leptonic parameters when ``additional chiral
coupling'' \newline is small}

In Table 1 for the $V+A$\ and for the $S\mp P$\ couplings, we list
the
``expanded forms'' of the above expressions for the case in which
there is a
single additional chiral coupling $(g_i/2\Lambda _i)$\ which is
small
relative to the standard $V-A$\ coupling $(g_L)$. Similarly, in
Table
2 is
listed the formulas for the additional tensorial couplings. The
tensorial
couplings include the sum and difference of the weak magnetism
and
electricity couplings, $g_{\pm }=f_M\pm f_E$, which involve the
momentum
difference $q_\rho =k_\tau -p_\nu $. The alternative tensorial
couplings $%
\tilde g_{\pm }={g^{+}}_{T^{+}\pm T_5^{+}}$ instead involve
$k_\tau +p_\nu $.%

Notice that except for the following coefficients the formulas
tablulated in
these two tables are short and simple. As above we usually suppress
the
entry in the ``$i$'' subscript on ``$\Lambda _i$.'' For \newline Table
1 these
coefficients are
\begin{equation}
\begin{array}{cc}
a=\frac{4m_\rho ^2}{m\Lambda }\frac{(1-\frac{m_\rho
^2}{m^2})}{(1-4\frac{%
m_\rho ^4}{m^4})} & d=\frac m{4\Lambda }(1-
\frac{m_\rho ^2}{m^2})\frac{(1-2\frac{m_\rho
^2}{m^2})}{(1+2\frac{m_\rho ^2}{%
m^2})} \\ b=\frac{m^2}{2\Lambda ^2}\frac{(1-\frac{m_\rho
^2}{m^2})^2}{(1-4%
\frac{m_\rho ^4}{m^4})} & e=
\frac{m^2}{4\Lambda ^2}\frac{(1-\frac{m_\rho
^2}{m^2})^2}{(1+2\frac{m_\rho ^2%
}{m^2})} \\ c=\frac{m_\rho ^2}{\Lambda ^2}\frac{(1-\frac{m_\rho
^2}{m^2})^2}{%
(1-4\frac{m_\rho ^4}{m^4})} & f=\frac m{2\Lambda }(1-
\frac{m_\rho ^2}{m^2})
\end{array}
\end{equation}
The coefficients for Table 2 are
\begin{equation}
\begin{array}{cc}
g=\frac{2m_\rho ^2}{m\Lambda }\frac{(1-4\frac{m_\rho
^2}{m^2})}{(1-4\frac{%
m_\rho ^4}{m^4})} & l=
\frac{m(1+9\frac{m_\rho ^2}{m^2}+2\frac{m_\rho
^4}{m^4})}{2\Lambda (1+2\frac{%
m_\rho ^2}{m^2})} \\ h=\frac{m_\rho ^2}{2\Lambda ^2}\frac{(1-
4\frac{m_\rho ^2%
}{m^2})}{(1-4\frac{m_\rho ^4}{m^4})} & m=
\frac{m_\rho ^2(2+\frac{m_\rho ^2}{m^2})}{2\Lambda
^2(1+2\frac{m_\rho ^2}{m^2%
})} \\ j=\frac{m_\rho ^2}{\Lambda ^2}\frac{(1-\frac{m_\rho
^2}{m^2})}{(1-4%
\frac{m_\rho ^4}{m^4})} & n=
\frac{m_\rho ^2(1-\frac{m_\rho ^2}{m^2})}{2\Lambda
^2(1+2\frac{m_\rho ^2}{m^2%
})} \\ k=\frac{3m_\rho ^2}{2\Lambda ^2(1+2\frac{m_\rho
^2}{m^2})} & o=\frac
m{2\Lambda }(1-\frac{m_\rho ^2}{m^2})
\end{array}
\end{equation}
Should experimental measurements indicate other than a pure
$g_L$
value of a semi-leptonic parameter, a smearing and more
sophisticated treated of these coefficients will be warrented.

Upon comparing the entries in these two tables, notice that (i) a
single
additional \ $\tilde g_{+}={g^{+}}_{T^{+}+T_5^{+}}\ $coupling
does not
change the values from that of the pure $V-A$\ coupling, and that
(ii) a
single additional \ $\tilde g_{-}={g^{+}}_{T^{+}-T_5^{+}}\
$coupling is
equivalent to a single additional $V+A$\ coupling, except for the
interpretation of their respective chirality parameters.  This follows
as a consequence of Eqs.(14, 15) and the absence of contributions
from the $S^-$ and $P^-$ couplings to the $\rho$, $a_1$, and
$K^*$
modes.

We have displayed this equivalence in Table 2 to emphasize the fact
that the commonly assumed total absence of $\tilde g_{\pm}$
couplings in tau lepton decays is supported by tests of the
experimental/theoretical normalization of the decay rates, such as
by
universality tests in lepton physics; however, this assumption is not
directly supported by the empirical values of other semi-leptonic
decay parameters.

\section{SPIN-CORRELATION FUNCTIONS IN TERMS OF
THE
SEMI-LEPTONIC PARAMETERS}

\subsection{The full S2SC function}

For the production decay sequence $e^{-}e^{+}\rightarrow
Z^o,\gamma
^{*}\rightarrow \tau ^{-}\tau ^{+}\rightarrow (\rho ^{-}\nu )(\rho
^{+}\bar
\nu )$\ followed by $\rho ^{ch}\rightarrow \pi ^{ch}\pi ^o$\ the full
``Stage 2 Spin-Correlation'' function (S2SC) including both $\nu
_L$,\ $\nu
_R\ $helicities and both   $\bar \nu _R$,\ $\bar \nu _L\ $helicities is
given by
\ber
\begin{array}{c}
{\bf I}_7={\bf I(}E_1,E_2,\phi ;\tilde \theta _{a,}\tilde \phi _a;\tilde
\theta _{b,}\tilde \phi _b{\bf )} \\ =\stackunder{h_1,h_2}{\sum
}\left|
T(h_1,h_2)\right| ^2\ {\bf R}_{h_1,h_1}\ {\bf \bar R}_{h_2,h_2}\
\\ +e^{\iota
\phi }T(++)T^{*}(--){\bf r}_{+-}{\bf \bar r}_{+-}+e^{-\iota \phi
}T(--)T^{*}(++){\bf r}_{-+}{\bf \bar r}_{-+}
\end{array}
\eer
where $T(\lambda _1,\lambda _2)\ $are the production helicity
amplitudes
given in Ref. \cite{11} which describe $Z^o,\gamma
^{*}\rightarrow \tau ^{-}\tau ^{+}.$   This formula also holds if
either, or both, $\tau ^{\pm }\rightarrow
a_{_1}^{\pm }\nu \ $followed by $a_{_1}^{\pm }\rightarrow (3\pi
)^{\pm }.$
The specific $\tau ^{\mp }\ $decay channel determines which
``composite
decay density matrix'' ${\bf R}_{h_1,h_1}$,\ or $\ {\bf \bar
R}_{h_2,h_2}$\
, is to be inserted.

The literature on polarimetry methods and spin-correlation function
in tau physics includes Refs. \cite{a2, c1, C94, C94a, ch2}.

{\bf Formulas for }$\tau \rightarrow \rho \nu $:

Including both $\nu _L\ $and $\nu _R\ $helicities and using a
``compact
boldface formalism,'' we find \cite{C94a} the composite decay
density matrix for $\tau
^{-}\rightarrow \rho ^{-}\nu \rightarrow (\pi ^{-}\pi ^o)\nu $ is%
\ber
{\bf R=}\left(
\begin{array}{cc}
{\bf R}_{++} & e^{\iota \phi _1^\tau }
{\bf r}_{+-} \\ e^{-\iota \phi _1^\tau }{\bf r}_{-+} & {\bf R}_{--}
\end{array}
\right)
\eer
In terms of the semi-leptonic parameters, the diagonal elements  are
\beq
{\bf R}_{\pm \pm }={\bf n}_a[1\pm {\bf f}_a\cos \theta _1^\tau
]\mp (1/\sqrt{%
2})\sin \theta _1^\tau \sin 2\tilde \theta _a\ {\cal R}_\rho [\omega
\cos \tilde \phi
_a+\eta ^{\prime }\sin \tilde \phi _a]
\eeq
These give the angular distributions $ \frac{dN}{d(\cos \theta
_1^\tau )d(\cos \tilde \theta _a)d \tilde \phi_a} $ for the polarized
$\tau^{-}$ decay
chain, see Eq.(1) above.  The off-diagonal elements depend on
\ber
\begin{array}{c}
{\bf r}_{+-}=({\bf r}_{-+})^{*} \\ ={\bf n}_a{\bf f}_a\sin \theta
_1^\tau
+(1/\sqrt{2})\sin 2\tilde \theta _a\ {\cal R}_\rho {\cos \theta _1^\tau
[\omega \cos \tilde
\phi _a+\eta ^{\prime }\sin \tilde \phi _a]+\iota [\omega \sin \tilde
\phi
_a-\eta ^{\prime }\cos \tilde \phi _a]\ }
\end{array}
\eer
In Eqs.(40, 41),%
\ber
\left(
\begin{array}{c}
{\bf n}_a \\ {\bf n}_a{\bf f}_a
\end{array}
\right) =\cos ^2\tilde \theta _a\frac{\Gamma _L^{\pm }}{\Gamma
}\pm
\frac 12\sin ^2\tilde \theta _a\frac{\Gamma _T^{\pm }}{\Gamma }
\eer
or equivalently%
\ber
\begin{array}{c}
{\bf n}_a=\frac 1{8}(3+\cos 2\tilde \theta _a+\sigma {\cal S}_\rho
[1+3\cos 2\tilde
\theta _a]) \\ {\bf n}_a{\bf f}_a=\frac 1{8}(\xi [1+3\cos 2\tilde
\theta _a]+\zeta {\cal S}_\rho [3+\cos 2\tilde \theta _a])
\end{array}
\eer

Similarly, for the conjugate process  $\tau ^{+}\rightarrow \rho
^{+}\bar
\nu \rightarrow (\pi ^{+}\pi ^o)\bar \nu \ $including both $\bar \nu
_R\ $and $%
\bar \nu _L\ $helicities
\ber
{\bf \bar R=}\left(
\begin{array}{cc}
{\bf \bar R}_{++} & e^{\iota \phi _2^\tau }
{\bf \bar r}_{+-} \\ e^{-\iota \phi _2^\tau }{\bf \bar r}_{-+} & {\bf
\bar R}%
_{--}
\end{array}
\right)
\eer
In terms of the semi-leptonic parameters, the diagonal elements are
\beq
{\bf \bar R}_{\pm \pm }={\bf n}_b[1\mp {\bf f}_b\cos \theta
_2^\tau ]\pm (1/%
\sqrt{2})\sin \theta _2^\tau \sin 2\tilde \theta _b\ {\cal R}_\rho [\bar
\omega \cos \tilde
\phi _b-\bar \eta ^{\prime }\sin \tilde \phi _b]
\eeq
and%
\ber
\begin{array}{c}
{\bf \bar r}_{+-}=({\bf \bar r}_{-+})^{*} \\ =-{\bf n}_b{\bf
f}_b\sin
\theta
_2^\tau -(1/\sqrt{2})\sin 2\tilde \theta _b\ {\cal R}_\rho {\cos \theta
_2^\tau [\bar
\omega \cos \tilde \phi _b-\bar \eta ^{\prime }\sin \tilde \phi
_b]+\iota
[\bar \omega \sin \tilde \phi _b+\bar \eta ^{\prime }\cos \tilde \phi
_b]\}
\end{array}
\eer
In Eqs.(45, 46),%
\ber
\left(
\begin{array}{c}
{\bf n}_b \\ {\bf n}_b{\bf f}_b
\end{array}
\right) =\cos ^2\tilde \theta _b\frac{\bar \Gamma _L^{\pm }}{\bar
\Gamma }%
\pm \frac 12\sin ^2\tilde \theta _b\frac{\bar \Gamma _T^{\pm
}}{\bar
\Gamma }
\eer
or equivalently%
\ber
\begin{array}{c}
{\bf n}_b=\frac 1{8}(3+\cos 2\tilde \theta _b+\bar \sigma {\cal
S}_\rho [1+3\cos
2\tilde \theta _b]) \\ {\bf n}_b{\bf f}_b=\frac 1{8}(\bar \xi [1+3\cos
2\tilde \theta _b]+\bar \zeta {\cal S}_\rho [3+\cos 2\tilde \theta _b])
\end{array}
\eer

{\bf Formulas for }$\tau \rightarrow a_1\nu $:

For the kinematic description of $\tau ^{-}\rightarrow a_1^{-}\nu
\rightarrow (\pi _1^{-}\pi _2^{-}\pi _3^{+})\nu $\ , the normal to
the
$(\pi
_1^{-}\pi _2^{-}\pi _3^{+})\ $decay triangle is used in place of the
$\pi
^{-}$\ momentum direction of the $\tau ^{-}\rightarrow \rho ^{-
}\nu
\rightarrow (\pi ^{-}\pi ^o)\nu \ $sequential decay \cite{bj}.

Including both $\nu _L\ $and $\nu _R\ $helicities, we find the
composite
decay density matrix for $\tau ^{-}\rightarrow a_1^{-}\nu
\rightarrow (\pi
_1^{-}\pi _2^{-}\pi _3^{+})\nu \ $ is%
\beq
{\bf R}^\nu =S_1^{+}{\bf R}^{+}+S_1^{-}{\bf R}^{-}
\eeq
where ${\bf R}^{\pm }$\ have the same the same form as the
earlier
matrix,
Eq.(39), except the elements now also have ``$\pm $'' superscripts,
see
below.\ $S_1^{\pm }\ $depend on the strong-interaction form-
factors
used to
describe the decay $a_1^{-}\rightarrow \pi _1^{-}\pi _2^{-}\pi
_3^{+}$.\
However, when the 3-body Dalitz plot is integrated over, only the
$S_1^{+}\ $%
term remains, so it can be absorbed into the overall normalization
factor
which removes any arbitrary form-factor dependence. In Eq.(49),
the
${\bf R}%
^{+}\ $composite decay matrix elements are%
\ber
\begin{array}{c}
{\bf R}_{\pm \pm }^{+}=\{Eq.(40)\ with\ (1/\sqrt{2})\rightarrow (-
1/\sqrt{2}%
)\} \\ {\bf r}_{+-}^{+}=({\bf r}_{-+}^{+})^{*} \\ =\{Eq.(41)\ with\
(1/\sqrt{2}%
)\rightarrow (-1/\sqrt{2})\}
\end{array}
\eer
with%
\ber
\left(
\begin{array}{c}
{\bf n}_a \\ {\bf n}_a{\bf f}_a
\end{array}
\right) =\sin ^2\tilde \theta _a\frac{\Gamma _L^{\pm }}{\Gamma
}\pm
(1-\frac 12\sin ^2\tilde \theta _a)\frac{\Gamma _T^{\pm
}}{\Gamma
}
\eer
or equivalently%
\ber
\begin{array}{c}
{\bf n}_a=\frac 1{16}(10-2\cos 2\tilde \theta _a-\sigma {\cal
S}_\rho
[5+3\cos 2\tilde
\theta _a]) \\ {\bf n}_a{\bf f}_a=\frac 1{16}(-\xi [5+3\cos 2\tilde
\theta _a]+\zeta {\cal S}_\rho [10-2\cos 2\tilde \theta _a])
\end{array}
\eer

Similarly, the ${\bf R}^{-}\ $composite decay matrix elements are
\beq
{\bf R}_{\pm \pm }^{-}=-{\bf n}_a^{-}[1\mp {\bf f}_a^{-}\cos
\theta _1^\tau
]\mp (\sqrt{2})\sin \theta _1^\tau \sin \tilde \theta _a\ {\cal R}_\rho
[\eta \cos \tilde
\phi _a+\omega ^{\prime }\sin \tilde \phi _a]
\eeq
with%
\ber
\left(
\begin{array}{c}
{\bf n}_a^{-} \\ {\bf n}_a^{-}{\bf f}_a^{-}
\end{array}
\right) =\cos \tilde \theta _a\frac{\Gamma _T^{\mp }}{\Gamma }
\eer
or%
\ber
\begin{array}{c}
{\bf n}_a^{-}=\frac 12\cos \tilde \theta _a[\xi -\zeta {\cal S}_\rho ]
\\
{\bf n}_a^{-}%
{\bf f}_a^{-}=\frac 12\cos \tilde \theta _a[1-\sigma {\cal S}_\rho ]
\end{array}
\eer
Also%
\ber
\begin{array}{c}
{\bf r}_{+-}^{-}=({\bf r}_{-+}^{-})^{*} \\ =\frac 12\sin \theta
_1^\tau \cos
\tilde \theta _a[1-\sigma {\cal S}_\rho ]+\sqrt{2}\sin \tilde \theta _a\
{\cal R}_\rho {\cos \theta
 _1^\tau [\eta \cos \tilde \phi _a+\omega ^{\prime }\sin \tilde \phi
_a]+\iota [\eta \sin \tilde \phi _a-\omega ^{\prime }\cos \tilde \phi
_a]\}
\end{array}
\eer

For the conjugate process  $\tau ^{+}\rightarrow a_1{}^{+}\bar \nu
\rightarrow (\pi _1^{+}\pi _2^{+}\pi _3^o)\bar \nu $\ ,%
\beq
{\bf \bar R}^{\bar \nu }=\bar S_1^{+}{\bf \bar R}^{+}+\bar
S_1^{-
}{\bf \bar R%
}^{-}
\eeq
The  ${\bf \bar R}^{+}\ $matrix elements are%
\ber
\begin{array}{c}
{\bf \bar R}_{\pm \pm }^{+}=\{Eq.(45)\ with\
(1/\sqrt{2})\rightarrow (-1/\sqrt{%
2})\} \\ {\bf \bar r}_{+-}^{+}=({\bf \bar r}_{-+}^{+})^{*} \\
=\{Eq.(46)\
with\ (1/\sqrt{2})\rightarrow (-1/\sqrt{2})\}
\end{array}
\eer
with%
\ber
\left(
\begin{array}{c}
{\bf n}_b \\ {\bf n}_b{\bf f}_b
\end{array}
\right) =\sin ^2\tilde \theta _b\frac{\bar \Gamma _L^{\pm }}{\bar
\Gamma }%
\pm (1-\frac 12\sin ^2\tilde \theta _b)\frac{\bar \Gamma _T^{\pm
}}{\bar
\Gamma }
\eer
or%
\ber
\begin{array}{c}
{\bf n}_b=\frac 1{16}(10-2\cos 2\tilde \theta _b-\bar \sigma {\cal
S}_\rho [5+3\cos
2\tilde \theta _b]) \\ {\bf n}_b{\bf f}_b=\frac 1{16}(-\bar \xi
[5+3\cos
2\tilde \theta _b]+\bar \zeta {\cal S}_\rho [10-2\cos 2\tilde \theta
_b])
\end{array}
\eer

The ${\bf \bar R}^{-}\ $matrix elements are
\beq
{\bf \bar R}_{\pm \pm }={\bf n}_b^{-}[1\pm {\bf f}_b^{-}\cos
\theta _2^\tau
]\mp \sqrt{2}\sin \theta _2^\tau \sin \tilde \theta _b\ {\cal R}_\rho
[\bar \eta \cos \tilde
\phi _b-\bar \omega ^{\prime }\sin \tilde \phi _b]
\eeq
and%
\ber
\begin{array}{c}
{\bf \bar r}_{+-}^{-}=({\bf \bar r}_{-+}^{-})^{*} \\ =\frac 12\sin
\theta
_2^\tau \cos \tilde \theta _b[1-\bar \sigma {\cal S}_\rho
]+\sqrt{2}\sin \tilde \theta
_b\ {\cal R}_\rho {\cos \theta _2^\tau [\bar \eta \cos \tilde \phi _b-
\bar \omega ^{\prime
}\sin \tilde \phi _b]+\iota [\bar \eta \sin \tilde \phi _b+\bar \omega
^{\prime }\cos \tilde \phi _b]\}
\end{array}
\eer
with%
\ber
\left(
\begin{array}{c}
{\bf n}_b^{-} \\ {\bf n}_b^{-}{\bf f}_b^{-}
\end{array}
\right) =\cos \tilde \theta _b\frac{\bar \Gamma _T^{\mp }}{\bar
\Gamma }
\eer
or%
\ber
\begin{array}{c}
{\bf n}_b^{-}=\frac 12\cos \tilde \theta _b[\bar \xi -\bar \zeta {\cal
S}_\rho ] \\ {\bf n}%
_b^{-}{\bf f}_b^{-}=\frac 12\cos \tilde \theta _b[1-\bar \sigma {\cal
S}_\rho ]
\end{array}
\eer

\subsection{The simplest S2SC function}

The simpler 4 variable S2SC function including both $\nu \ $and
both $\bar
\nu $\ helicities is
\ber
\begin{array}{c}
{\bf I}_4={\bf I(}E_1,E_2,\tilde \theta _1,\tilde \theta _2{\bf )} \\
=\left| T(+,-)\right| ^2 {\bf \rho }_{++} {\bf \bar \rho
}_{--}+\left|
T(-,+)\right| ^2 {\bf \rho }_{--} {\bf \bar \rho
}_{++}+\left| T(+,+)\right| ^2%
{\bf \rho }_{++} {\bf \bar \rho }_{++}+\left| T(-,-
)\right| ^2 {\bf \rho }_{--}%
{\bf \bar \rho }_{--}
\end{array}
\eer
This formula is in terms of the {\it integrated} composite decay
density
matrices for the $\tau ^{\pm }\rightarrow \rho ^{\pm }\nu \ $and/or
for the $%
\tau ^{\pm }\rightarrow a_1^{\pm }\nu $ decay chains with $\rho
^{\pm
}\rightarrow (2\pi )^{\pm }$\ and  a$_1^{\pm }\rightarrow (3\pi
)^{\pm }$\ .  Note that as for the $\bf R$'s in the preceding section,
here in Eq.(65) the $\rho$'s include both neutrino helicities.  Here,
for convenience we suppress their ``boldface font''.

{\bf Formulas for }$\tau \rightarrow \rho \nu :$

For $\tau ^{-}\rightarrow \rho ^{-}\nu \rightarrow (\pi ^{-}\pi
^o)\nu
$,
with $\tau ^{-}$\ helicity $\lambda _1=h/2$%
\ber
\begin{array}{c}
{\bf \rho }_{hh}\equiv \frac 1{\Gamma }\frac{dN}{d(\cos \theta
_1^\tau )d(\cos \tilde \theta _1)} \\ =\frac 18(3+\cos 2\tilde \theta
_1)S+\frac 1{16}(1+3\cos 2\tilde \theta _1)D
\end{array}
\eer
where
\beq
S=1+h\zeta {\cal S}_\rho  \cos \theta _1^\tau
\eeq
\beq
D=-S(1-\cos 2\omega _1)+(\sigma {\cal S}_\rho +h\xi \cos \theta
_1^\tau )(1+3\cos 2\omega
_1)+h\omega {\cal R}_\rho 4\sqrt{2}\sin 2\omega _1\sin \theta
_1^\tau .
\eeq
Formulas for the Wigner rotation angles $\omega_{1,2}$ which are
solely functions respectively of $E_{1,2}$ are given in \cite{C94}.

It is important to note that if $\tilde \theta _1$\ is integrated out, i.e.
if the polarimetry information from the $\rho ^{-}\rightarrow (2\pi
)^{-}\ $%
stage is not included, then $D$  doesn't contribute.  In this manner,
$\zeta $ is
measurable. Then inclusion of the $\tilde \theta _1$\  dependence
gives $D$ and
also enables separation of  $\xi $ and $\omega $ because of their
differing
dependence on $\tilde \theta _1$%
{}.

For the CP conjugate process $\tau ^{+}\rightarrow \rho ^{+}\bar
\nu
\rightarrow (\pi ^{+}\pi ^o)\bar \nu $, with $\tau ^{+}$\ helicity
$\lambda
_1=h/2$
\ber
\begin{array}{c}
{\bf \bar \rho }_{hh}\equiv \frac 1{\tilde \Gamma
}\frac{dN}{d(\cos
\theta
_2^\tau )d(\cos \tilde \theta _2)} \\ =\frac 18(3+\cos 2\tilde \theta
_2)\bar S+\frac 1{16}(1+3\cos 2\tilde \theta _2)\bar D
\end{array}
\eer
where
\beq
\bar S=1-h\bar \zeta {\cal S}_\rho \cos \theta _2^\tau
\eeq
\beq
\bar D=-\bar S(1-\cos 2\omega _2)+(\bar \sigma {\cal S}_\rho -h\bar
\xi \cos \theta
_2^\tau )(1+3\cos 2\omega _2)-h\bar \omega {\cal R}_\rho
4\sqrt{2}\sin 2\omega _2\sin
\theta _2^\tau .
\eeq

{\bf Formulas for }$\tau \rightarrow a_1\nu :$

For $\tau ^{-}\rightarrow a_1^{-}\nu \rightarrow (3\pi )^{-}\nu $,
with $%
\tau ^{-}$\ helicity $\lambda _1=h/2$

where%
\ber
\begin{array}{c}
{\bf \rho }_{hh}\equiv \frac 1{\Gamma }\frac{dN}{d(\cos \theta
_1^\tau )d(\cos \tilde \theta _1)} \\ =\frac 14(3+\cos 2\tilde \theta
_1)S_{a_1}-\frac 1{32}(1+3\cos 2\tilde \theta _1)D_{a_1}
\end{array}
\eer
\beq
S_{a_1}=1+h\zeta {\cal S}_\rho \cos \theta _1^\tau
\eeq
\beq
D_{a_1}=S_{a_1}(3+\cos 2\omega _1)+(\sigma {\cal S}_\rho +h\xi
\cos \theta _1^\tau
)(1+3\cos 2\omega _1)+h\omega {\cal R}_\rho 4\sqrt{2}\sin
2\omega _1\sin \theta _1^\tau .
\eeq
The remarks above, following the analogous formulas in the $\rho $
case,
also apply here.

For the CP conjugate process $\tau ^{+}\rightarrow a_1^{+}\bar
\nu
\rightarrow (3\pi )^{+}\bar \nu $, with $\tau ^{+}$\ helicity
$\lambda _2=h/2
$
\ber
\begin{array}{c}
{\bf \bar \rho }_{hh}\equiv \frac 1{\bar \Gamma }\frac{dN}{d(\cos
\theta
_2^\tau )d(\cos \tilde \theta _2)} \\ =\frac 14(3+\cos 2\tilde \theta
_2)\bar S_{a_1}-\frac 1{32}(1+3\cos 2\tilde \theta _2)\bar D_{a_1}
\end{array}
\eer
where
\beq
\bar S_{a_1}=1-h\bar \zeta {\cal S}_\rho \cos \theta _2^\tau
\eeq
\beq
\bar D_{a_1}=\bar S_{a_1}(3+\cos 2\omega _2)+(\bar \sigma {\cal
S}_\rho -h\bar \xi \cos
\theta _2^\tau )(1+3\cos 2\omega _2)-h\bar \omega {\cal R}_\rho
4\sqrt{2}\sin 2\omega
_2\sin \theta _2^\tau .
\eeq

\section{Tests for non-CKM-type leptonic CP violation}

By CP invariance each of the barred semi-leptonic parameters
should
equal, within experimental errors, its unbarred associate.  However,
as was shown in Ref. \cite{C94}, if only $\nu_L$ and $\bar\nu_R$
exist,  there are two simple tests for ``non-CKM-type" leptonic CP
violation in $\tau \rightarrow \rho \nu$ decay.  Normally a CKM
leptonic-phase will contribute equally at tree level to both the
$\tau^-
$ and $\tau^+$ decay amplitudes (for exceptions see footnotes 14,
15 in Ref. \cite{C94}).  These two tests follow because by CP
invariance $|B\left( \lambda_{\bar\rho},\lambda_{\bar\nu}\right) | =
|A\left(-\lambda_{\bar\rho},-\lambda_{\bar\nu}\right) |$.  So the
two
tests for leptonic CP
violation are: %
\beq
\beta _a=\beta _b \hspace{2pc} {\bf first \hspace*{.4pc} test}
\eeq
where $\beta _a=\phi _{-1}^a-\phi _0^a$, $\beta _b=\phi _1^b-\phi
_0^b$, and
\beq
r_a=r_b \hspace{2pc} {\bf second  \hspace*{.4pc} test}
\eeq
where%
\beq
r_a=\frac{|A\left( -1,-\frac 12\right) |}{|A\left( 0,-\frac 12\right)
|},r_b=%
\frac{|B\left( 1,\frac 12\right) |}{|B\left( 0,\frac 12\right) |}
\eeq
For sensitivity levels for $\tau \rightarrow \rho \nu$ decay, see Ref.
\cite{1}.

This analysis can be easily generalized \cite{C94a} to the $\tau
\rightarrow a_1\nu $ decay mode in which the
$a_1$ has
the opposite $CP$ quantum number to that of the $\rho $ : For the
$\tau ^{-}\rightarrow a_1^{-}\nu \rightarrow \left( \pi ^{-}\pi
^{-}\pi ^{+}\right) \nu ,\left( \pi ^o\pi ^o\pi ^{-}\right) \nu $ modes,
the composite-decay-density matrix is given by

\begin{equation}
\begin{array}{c}
\rho _{hh}=
 \left( 1+h\cos \theta _1^\tau \right) \left[ \sin ^2\omega _1\cos
^2\tilde \theta _1 + ( 1- \frac 12\sin ^2\omega _1 ) \sin ^2\tilde
\theta
_1 \right] \\
+ \frac{r_a^2}2\left( 1-h\cos \theta _1^\tau \right) \left[ \left( 1+\cos
^2\omega_1\right) \cos ^2\tilde \theta _1  +\left( 1+\frac 12\sin
^2\omega
_1\right) \sin^2\tilde \theta _1 ] \\
-h\frac{r_a}{\sqrt{2}}\cos \beta _a\sin \theta
_1^\tau \sin 2\omega _1\left[ \cos ^2\tilde \theta _1-\frac
12\sin^2\tilde\theta_1\right]
\end{array}
\end{equation}

Table 3 shows that the sensitivity of the $a_1$ mode, versus that of
the $\rho$ mode, is about 2 times better for the $r_a$ measurement
and is about 5 times worse for the $\beta$ measurements.  The
simpler $I_4$ function was used for $\sigma \left( r_a \right)$ and
the full $I_7$ was used for the other $\sigma$'s.  The $CP$ and
$CP\tilde T_{FS}$ predictions for the phase relation between
$\beta
_a$ and $\beta _b$ are opposite, see Table 3 in \cite{e4}, so this
provides a method for distinguishing between a new physics effect
due to an unusual $CP$-violating final state interaction and one
with
a different mechanism of $CP$ violation.

It is also easy to generalize these simple tests so as to also include
$\nu
_R$ and $\bar
\nu
_L$ couplings.  The necessary 4-variable S2SC is given by

\begin{equation}
\begin{array}{c}
I\left( E_\rho ,E_{\bar \rho },\tilde \theta _1,\tilde \theta _2\right)
\mid_{\nu _R,\bar \nu _L}=I_4 +\left( \lambda _R\right) ^2I_4\left(
\rho \rightarrow \rho ^R\right)
+\left( \bar \lambda _L\right) ^2I_4\left( \bar \rho \rightarrow
\bar\rho
^L\right) \\
+\left( \lambda _R\bar \lambda _L\right) ^2I_4\left( \rho \rightarrow
\rho^R,\bar \rho \rightarrow \bar \rho ^L\right)
\end{array}
\end{equation}
where $\lambda _R\equiv $ $\frac{|A\left( 0,\frac 12\right)
|}{|A\left(
0,-\frac 12\right) |},$ $\bar \lambda _L\equiv $ $\frac{|B\left( 0,-
\frac
12\right) |}{|B\left( 0,\frac 12\right) |}$ give the moduli's of the $\nu
_R$
and $\bar \nu _L$ amplitudes versus the standard amplitudes. The
corresponding
composite density matrices for $\tau \rightarrow \rho \nu $ with
$\nu _R$
and $\bar \nu _L$ final state particles are given by the substitution
rules:%
\ber
\rho _{hh}^R=\rho _{-h,-h}\left( r_a\rightarrow r_a^R,\beta
_a\rightarrow
\beta _a^R\right)  \\
\bar \rho _{hh}^L=\bar \rho _{-h,-h}\left( r_b\rightarrow
r_b^L,\beta
_b\rightarrow \beta _b^L\right)
\eer
where the  $\nu _R$ and $\bar \nu _L$ moduli ratios and phase
differences
are defined by $r_a^R\equiv $ $\frac{|A\left( 1,\frac 12\right) |}{%
|A\left( 0,\frac 12\right) |},$ $r_b^L\equiv $ $\frac{|B\left( -1,-\frac
12\right) |}{|B\left( 0,-\frac 12\right) |},\beta _a^R\equiv \phi _1^a-
\phi
_0^{aR},\beta _b^L\equiv \phi _{-1}^b-\phi _0^{bL}$.  The two
addtitional tests for ``non-CKM-type" leptonic CP violation if R-
handed $\nu$ and L-handed $\bar\nu$ exist are
\beq
{\beta _a}^R={\beta _b}^L \hspace{2pc} {\bf first \hspace*{.4pc}
\nu_R / {\bar\nu_L} \hspace*{.4pc} test}
\eeq
\beq
{r_a}^R={r_b}^L \hspace{2pc} {\bf second \hspace*{.4pc} \nu_R
/
{\bar\nu_L}  \hspace*{.4pc} test}
\eeq

\section{DESCRIPTION OF $\tau ^{-}\rightarrow \pi ^{-}\nu
,K^{-
}\nu $}

The only observables for each of the $\tau ^{-}\rightarrow \pi ^{-
}\nu
,K^{-}\nu $\ modes which can be measured by spin-correlations are
the
chirality parameter $\xi _\pi =\frac{\left| A(-\frac 12)\right| ^2-\left|
A(\frac 12)\right| ^2}{\left| A(-\frac 12)\right| ^2+\left| A(\frac
12)\right| ^2}$ and the $\Gamma (\tau ^{-}\rightarrow \pi ^{-}\nu
)$,
\newline or $%
\Gamma (\tau ^{-}\rightarrow K^{-}\nu )$, partial width. The
relative phase
of the $A(\lambda _\nu )=A(\mp \frac 12)$\ amplitudes can not be
measured
unless, e.g. the $\nu _L$and $\nu _R$ have a common final decay
channel. For
$\tau ^{-}\rightarrow \pi ^{-}\nu ,$or $K^{-}\nu $,\ the $(k_\tau
+p_\nu )$\
effective couplings $(k_\tau +p_\nu )_\alpha \ V_{\nu \tau }^\alpha
$\ and $%
(k_\tau +p_\nu )_\alpha \ A_{\nu \tau }^\alpha $ are equivalent to
the
standard $q_{\pi ,\alpha }\ V_{\nu \tau }^\alpha $\ and $q_{\pi
,\alpha }\
A_{\nu \tau }^\alpha $\ couplings. Here $V_{\nu \tau }^\alpha $\
and $A_{\nu
\tau }^\alpha $\ are as in Eqs.(11-13). The $S^{-}$\ and $P^{-}$\
couplings can
contribute to the $\pi ^{-\text{\ }}$and $K^{-}$\ channels, whereas
they do
not for the $\rho ,a_{1,}K^{*}$\ modes. However, since $q\cdot
V\sim \frac{%
m_\pi ^2}{2\Lambda }g_{S^{-}}$\ and $q\cdot A\sim \frac{m_\pi
^2}{2\Lambda }%
g_{P^{-}}$\ their contribution is strongly suppressed for $\Lambda
>(\sim
1GeV)$\ scales.

By Lorentz invariance, there are the equivalence theorems that
$S^{-
}\approx
S\approx T^{+}\approx V$\ and P$^{-}\approx P\approx
T_5^{+}\approx A$\ .
The general helicity amplitudes for $\tau ^{-}\rightarrow \pi ^{-}\nu
,$or $%
K^{-}\nu $, for the above $q\cdot V$\ and $q\cdot A$\ couplings
are
\begin{equation}
\begin{array}{c}
A(\mp \frac 12)=g_L(E_\rho \pm q_\pi )
\sqrt{m_\tau (E_\nu \pm q_\pi )}+g_R(E_\rho \mp q_\pi
)\sqrt{m_\tau (E_\nu
\mp q_\pi )} \\ +(
\frac{m_\tau }{2\Lambda _i}%
)[g_{S+P}+g_{S-P}+(g_{S^{-}+P^{-}}+g_{S^{-}-P^{-
}})(\frac{m_\pi ^2}{m_\tau
^2-m_\nu ^2})]\{(E_\rho \pm q_\pi )\sqrt{m_\tau (E_\nu \pm q_\pi
)} \\
+(E_\rho \mp q_\pi )
\sqrt{m_\tau (E_\nu \mp q_\pi )}\} \\ +\tilde g_{+}(
\frac{m_\tau }{2\Lambda })\{(-1+\frac{m_\pi ^2}{m_\tau ^2-
m_\nu
^2})(E_\rho
\pm q_\pi )\sqrt{m_\tau (E_\nu \pm q_\pi )} \\ +(1+
\frac{m_\pi ^2}{m_\tau ^2-m_\nu ^2})(E_\rho \mp q_\pi
)\sqrt{m_\tau (E_\nu
\mp q_\pi )}\} \\ +\tilde g_{-}(
\frac{m_\tau }{2\Lambda })\{(1+\frac{m_\pi ^2}{m_\tau ^2-m_\nu
^2})(E_\rho
\pm q_\pi )\sqrt{m_\tau (E_\nu \pm q_\pi )} \\ +(-1+\frac{m_\pi
^2}{m_\tau
^2-m_\nu ^2})(E_\rho \mp q_\pi )\sqrt{m_\tau (E_\nu \mp q_\pi
)}\}
\end{array}
\end{equation}
The $\xi _\pi \ $ parameter can be measured by the stage-one
energy-correlation function $I(E_1^\pi ,E_2^\pi )$\ where $\rho
_{\pm \pm
}=1\pm \xi _\pi \cos \theta _1^\tau ,$ $\bar \rho _{\pm \pm }=1\mp
\bar \xi
_\pi \cos \theta _2^\tau $\ . From Eq(87) the effective $\lambda
=\left|
g_{eff}/g_L\right| ^2$\ value follows for
\begin{equation}
\frac{\Gamma (\tau \rightarrow \pi \nu _\tau )}{\Gamma (\pi
\rightarrow \mu
\nu _\mu )}=\frac \lambda 2\frac{m_\tau ^3}{m_\mu ^2m_\pi
}\left(
\frac{1-%
\frac{m_\pi ^2}{m_\tau ^2}}{1-\frac{m_\mu ^2}{m_\pi ^2}}\right)
^2
\end{equation}
For example,
$$
\begin{array}{cc}
\lambda _{S+P}=\left| 1+\frac{m_\tau }{2\Lambda
}\frac{g_{S+P}}{g_L}\right|
^2, & \lambda _{\tilde g_{+}}=\left| 1-\frac{m_\tau }{2\Lambda
}\frac{\tilde
g_{+}}{g_L}(1-\frac{m_\pi ^2}{m_\tau ^2})\right| ^2
\end{array}
$$

\begin{center}
{\bf Acknowledgments}
\end{center}
For helpful discussions, we thank experimentalists and theorists at
Cornell,
DESY, Valencia, and at the Montreux workshop. This work was
partially
supported by U.S. Dept. of Energy Contract No. DE-FG 02-
96ER40291.

\section*{Appendix: The helicity amplitudes in terms of the chiral
couplings}

In Sec. 2, the simple symmetry relations among the amplitudes are
possible
because of the Jacob-Wick  phase conventions that were built into
the helicity formalism \cite{5}.  In combining these amplitudes with
results from calculations of
similar amplitudes by diagramatic methods, care must be exercised
to insure that the same phase
conventions are being used (c.f. appendix in \cite{11}).

The helicity amplitudes for $\tau^{-
}\rightarrow \rho
^{-}\nu _{L,R}$ for both $(V\mp A)$ couplings and
$%
m_\nu $ arbitrary are for $\nu _L$ so $\lambda _\nu =-\frac 12$,%
\ber
A\left( 0,-\frac 12\right) & = & g_L
\frac{E_\rho +q_\rho }{m_\rho } \sqrt{m_\tau \left( E_\nu
+q_\rho
\right) } -g_R
\frac{E_\rho -q_\rho }{m_\rho } \sqrt{m_\tau \left( E_\nu -
q_\rho
\right) } \\ A\left( -1,-\frac 12\right) & = & g_L
\sqrt{2m_\tau \left( E_\nu +q_\rho \right) } -g_R\sqrt{2m_\tau \left(
E_\nu -q_\rho \right) }.
\eer
and for $\nu _R$ so $\lambda _\nu =\frac 12$,%
\ber
A\left( 0,\frac 12\right) & = & -g_L
\frac{E_\rho -q_\rho }{m_\rho } \sqrt{m_\tau \left( E_\nu -
q_\rho
\right) }  +g_R
\frac{E_\rho +q_\rho }{m_\rho } \sqrt{m_\tau \left( E_\nu
+q_\rho
\right) } \\ A\left( 1,\frac 12\right) & = & -g_L
\sqrt{2m_\tau \left( E_\nu -q_\rho \right) } +g_R\sqrt{2m_\tau \left(
E_\nu +q_\rho \right) }
\eer
Note that
$g_L,g_R$
denote the
`chirality' of the coupling and $\lambda _\nu =\mp \frac 12$ denote
the
handedness of $\nu _{L,R}$.  For $(S \pm P)$ couplings, the
additional contributions are
\ber
A(0,-\frac 12) & =g_{S+P}(
\frac{m_\tau }{2\Lambda })\frac{2q_\rho }{m_\rho }\sqrt{m_\tau
(E_\nu
+q_\rho )}  +g_{S-P}(\frac{m_\tau }{2\Lambda })\frac{2q_\rho
}{m_\rho }%
\sqrt{m_\tau (E_\nu -q_\rho )}, \quad
A(-1,-\frac 12) & =0
\eer
\ber
A(0,\frac 12) & =g_{S+P}(
\frac{m_\tau }{2\Lambda })\frac{2q_\rho }{m_\rho }\sqrt{m_\tau
(E_\nu
-q_\rho )}  +g_{S-P}(\frac{m_\tau }{2\Lambda })\frac{2q_\rho
}{m_\rho }%
\sqrt{m_\tau (E_\nu +q_\rho )}, \quad
A(1,\frac 12) & =0
\eer
The two types of tensorial couplings, $g_\pm = f_M \pm f_E$ and
$\tilde{g}_{\pm}={g^+}_{T^+ \pm T_5^+}$,  give the additional
contributions
\begin{eqnarray*}
A\left( 0,\mp\frac 12\right)         & = & \mp g_{+} (
\frac{m_\tau }{2\Lambda }) \left[
\frac{E_\rho \mp q_\rho }{m_\rho } \sqrt{m_\tau \left( E_\nu
\pm q_\rho
\right) } - \frac{m_\nu }{m_\tau }
\frac{E_\rho \mp q_\rho }{m_\rho } \sqrt{m_\tau \left( E_\nu \mp
q_\rho
\right) } \right] \\
                                             &   & \pm g_{-} (
\frac{m_\tau }{2\Lambda }) \left[
 - \frac{m_\nu }{m_\tau }
\frac{E_\rho \pm q_\rho }{m_\rho } \sqrt{m_\tau \left( E_\nu \pm
q_\rho
\right) } + \frac{E_\rho \pm q_\rho }{m_\rho } \sqrt{m_\tau \left(
E_\nu
\mp q_\rho
\right) } \right] \\
                                           &   & \mp \tilde g_{+} (
\frac{m_\tau }{2\Lambda }) \left[
\frac{E_\rho \pm q_\rho }{m_\rho } \sqrt{m_\tau \left( E_\nu
\pm q_\rho
\right) } + \frac{m_\nu }{m_\tau }
\frac{E_\rho \mp q_\rho }{m_\rho } \sqrt{m_\tau \left( E_\nu \mp
q_\rho
\right) } \right] \\
                                             &   & \pm \tilde g_{-} (
\frac{m_\tau }{2\Lambda }) \left[
  \frac{m_\nu }{m_\tau }
\frac{E_\rho \pm q_\rho }{m_\rho } \sqrt{m_\tau \left( E_\nu \pm
q_\rho
\right) } + \frac{E_\rho \mp q_\rho }{m_\rho } \sqrt{m_\tau \left(
E_\nu
\mp q_\rho
\right) } \right]
\end{eqnarray*}
\begin{eqnarray*}
A\left( \mp 1,\mp\frac 12\right) & = & \mp \sqrt{2} g_{+} (
\frac{m_\tau }{2\Lambda }) \left[
  \sqrt{m_\tau \left( E_\nu
\pm q_\rho
\right) } -  \frac{m_\nu }{m_\tau }
 \sqrt{m_\tau \left( E_\nu \mp
q_\rho
\right) } \right] \\
                                             &   & \pm \sqrt{2} g_{-} (
\frac{m_\tau }{2\Lambda }) \left[
 - \frac{m_\nu }{m_\tau }
 \sqrt{m_\tau \left( E_\nu \pm
q_\rho
\right) } +  \sqrt{m_\tau \left( E_\nu
\mp q_\rho
\right) } \right] \\
                                  &   & \mp \sqrt{2} \tilde g_{+} (
\frac{m_\tau }{2\Lambda }) \left[
 \sqrt{m_\tau \left( E_\nu
\pm q_\rho
\right) } + \frac{m_\nu }{m_\tau }
 \sqrt{m_\tau \left( E_\nu \mp
q_\rho
\right) } \right] \\
                                             &   & \pm \sqrt{2} \tilde g_{-} (
\frac{m_\tau }{2\Lambda }) \left[
  \frac{m_\nu }{m_\tau }
 \sqrt{m_\tau \left( E_\nu \pm
q_\rho
\right) } +  \sqrt{m_\tau \left( E_\nu
\mp q_\rho
\right) } \right]
\end{eqnarrray*}

\begin{thebibliography}{333}
\bibitem{0} For recent experimental measurements of tau couplings
see ARGUS collab., DESY 94-120; ALEPH collab., Phys. Lett. {\bf
B321}, 168(1994); M. Schmidtler, Nuc. Phys. B(Proc. Suppl.) {\bf
40}, 265(1995); J. Raab, ibid. 413; M. Davier, ibid. 395; R.
Stroynowski, ibid. p. 569.
\bibitem{1} C.A. Nelson, SUNY BING 10/1/94, to appear in Phys.
Lett. (1995).
\bibitem{C94} C.A. Nelson, H.S. Friedman, S.Goozovat, J.A.
Klein,
L.R. Kneller, W.J. Perry, and S.A. Ustin, Phys. Rev. {\bf D50},
4544(1994); C.A. Nelson, SUNY BING 7/19/92; in {\em Proc. of
the Second Workshop on Tau Lepton Physics}, K.K. Gan (ed),
World Sci., Singapore, 1993.
\bibitem{C94a} C.A. Nelson, M. Kim, and H.-C. Yang, SUNY
BING 5/27/94; ICHEP94\#0100 (Glasgow).
\bibitem{5} M. Jacob and G. Wick, Ann. Phys. (N.Y.) {\bf 7},
209(1959).
\bibitem{6} See for example, E. Braaten, S. Narison and A. Pich,
Nucl. Phys {\bf B373}, 581(1992);  F. Le Diberder and A. Pich,
Phys. Lett. {\bf B286}, 147(1992); S. Narison and A. Pich, Phys.
Lett {\bf B304}, 359(1993); and S. Narison Nuc. Phys. B(Proc.
Suppl.) {\bf40}, 47(1995).
\bibitem{7} C.A. Nelson, Phys. Rev. {\bf D40}, 123(1989); {\bf
D41}, 2327(1990)(E).
\bibitem{th}  S. Weinberg, in {\em
Unification of Elementary Forces and Gauge Theories}, D.B. Cline
and F.E. Mills(eds), Harwood Pub., London, 1978; G. 't Hooft,
THU-94/15.
\bibitem{d0} S.M. Barr and W. Marciano, in {\em CP Violation},
C. Jarlskog(ed), World Sci., Singapore, 1989; W. Bernreuther, U.
Low, J.P. Ma, and O. Nachtmann, Z. Phys. {\bf C43}, 117(1989); J.
Bernabeu, N. Rius, and A. Pich, Phys. Lett. {\bf B257},
219(1991);
S. Goozovat and C.A. Nelson, Phys. Rev. {\bf  D44}, 2818(1991);
J.A. Grifols and A. Mendez, Phys. Lett. {\bf B255},  611(1991);
and
R. Escribano and E. Masso, Phys. Lett. {\bf B301}, 419(1993);
UAB-FT-317.
\bibitem{e4} C.A. Nelson, Nuc. Phys. B(Proc. Suppl.) {\bf 40},
525(1995).
\bibitem{11} S. Goozovat and C.A. Nelson, Phys. Rev. {\bf D44},
2818(1991).
\bibitem{bj} S.M. Berman and M. Jacob, SLAC Report No. 43
(1965), unpublished; Phys. Rev. {\bf 139},   B1023(1965).
\bibitem{a2} Y.-S. Tsai, Phys. Rev. {\bf D4}, 2821(1971); S.Y. Pi
and A.I. Sanda, Ann. Phys. (N.Y.) {\bf 106}, 171(1977); J. Babson
and E. Ma, Phys. Rev. {\bf D26}, 2497 (1982);  J.H. Kuhn and F.
Wagner, Nuc. Phys. {\bf B236}, 16(1984);
A. Rouge, Z. Phys {\bf C48},  75(1990); M. Davier, L. Duflot, F.Le
Diberder, and A. Rouge, Phys. Lett. {\bf B306},  411(1993); K.
Hagiwara,
A.D. Martin, and D. Zeppenfeld, Phys. Lett. {\bf  B235},
198(1990); B.K.
Bullock, K. Hagiwara, and A.D. Martin, Nuc. Phys. {\bf B359},
499
(1993).
\bibitem{c1} C.A. Nelson, Phys. Rev. Lett. {\bf 62},  1347(1989);
Phys.
Rev. {\bf D40},   123(1989); {\bf D41},  2327(E)(1990); Phys.
Rev.
{\bf D41},
2805(1990) ; W. Fetscher, Phys. Rev. {\bf D42},  1544(1990); H.
Thurn and H. Kolanoski, Z. Phys. {\bf C60},  277(1993).
\bibitem{ch2} Y.-S. Tsai, Phys. Rev. {\bf D51}, 3172(1995); talk
to
appear in {\em Proc. of Workshop on a Tau Charm Factory in Era
of
CESR/B Factory} SLAC, Aug. '94; Y.-S. Tsai, SLAC-PUB-95-
6916.

\newpage

\begin{center}
{\bf Table Captions}
\end{center}

Table 1: Semi-leptonic decay parameters for $\tau^{-}\rightarrow
\rho^{-}\nu, \ldots$ in the case of a single additional chiral coupling
($g_{\iota}$) which is small relative to the standard $V-A$
coupling
($g_L$).  This table is for the $V+A$ and for the $S \pm P$
couplings. The next table is for additional tensorial couplings. The
first 3 parameters give the partial widths, $\bf{\Gamma}$, for all
the
possible $\rho_{L,T} \nu_{L,R}$ final-state combinations; the
latter
4 parameters give the complete $\rho_L$ -$\rho_T$ interference
intensities, $\bf{\Gamma_{LT}}$.  In this paper $\cal RE$ ( $\cal
IM$ ) denote respectively the real (imaginary) parts of the quantity
inside the parentheses.

Table 2: Same as previous table except this table is for additional
tensorial couplings.
Here $g_\pm = f_M \pm f_E$ involves $ k_\tau - p_\nu $ whereas
$\tilde{g}_{\pm}={g^+}_{T^+ \pm T_5^+}$ involves $ k_\tau +
p_\nu $, see Eqs.(12,13).

Table 3: The ideal statistical errors for the two tests for ``non-
CKM-
type'' leptonic CP violation in $\tau^{-} \rightarrow a^{-}_{1}
\nu_L
$ decay, including both $a^{-}_{1} \rightarrow (2 \pi^{-} \pi^{+}
)$
and $(2 \pi^{o} \pi^{-} )$ modes. The $CP\~{T_{FS}}, CP,
\~{T_{FS}}$ labels denote the symmetries which would
respectively be violated if $r_a \neq r_b$, $\tilde\beta \neq 0$, etc.
Note that $\tilde\beta \equiv \beta_a - \beta_b$ and $\beta' \equiv
\beta_a + \beta_b$.  At $M_Z$ we assume $10^7$ Z-bosons and
assume $10^7$ $\tau^- \tau^+$ pairs at each of the other center-of-
mass energies.

\newpage

\begin{table*}[h]
\setlength{\tabcolsep}{2.pc}
\newlength{\digitwidth} \settowidth{\digitwidth}{\rm 0}
\catcode`?=\active \def?{\kern\digitwidth}
\caption{}
\label{tab1}
\begin{tabular*}{\textwidth}{@{}lllll}
\hline
                 & \multicolumn{2}{l}{$V \pm A
$ }
                 & \multicolumn{2}{l}{Additional $S \pm
P
$ }
\\
\cline{2-3} \cline{4-5}
                 & \multicolumn{1}{l}{Pure $g_L$}
                 & \multicolumn{1}{l}{Plus $g_R$}
                 & \multicolumn{1}{l}{Plus $g_{S+P}$}
                 & \multicolumn{1}{l}{Plus $g_{S-P}$}
\\
\hline
{$\bf  \; \; \Gamma$'s}                    &
&

&

&
                $
$
\\
$\zeta$                                          & $ 1
$ & $  \xi $

&
$1+a \frac{{\cal RE}(g_L^{*}g_{S+P})}{\left| g_L\right|
^2}  $ &
 $                      1-b \frac{\left| g_{S-
P}}{g_L\right| ^2}$
\\
$\sigma$                                       & $ 1  $
&  $ 1          $

&
$ \zeta    $
&
$1+c \frac{\left| g_{S-P} }{g_L\right| ^2}$
\\
$\xi$                                             & $
1$                 &  $\frac{\left|
g_L\right| ^2-\left| g_R\right| ^2}{\left| g_L\right|
^2+\left|g_R\right|^2}$&
 $  1$
&
$   1  $
\\
{$\bf  \; \; \Gamma_{LT}$'s }       &
&

&

&
              $
$
\\
$\omega$
& $ 1          & $          \xi$

&
$1-d \frac{{\cal RE}(g_L^{*}g_{S+P})}{ \left| g_L\right|
^2}$
&
 $1-e \frac{\left| g_{S-P}}{g_L\right| ^2}$
\\
$\eta$                                          & $  1 $
&  $  1   $

&
$\omega$
&
$\omega $
\\
$\omega'$                                   & $ 0 $
&  $ 0 $

&
$-f  \frac{{\cal IM}(g_L^{*}g_{S+P})}{\left|
g_L\right|^2}$
&
$  0 $
\\
$\eta'$                                        & $ 0 $
&  $  0 $

&
$
\omega'  $
&
$   0 $
\\
\hline
\multicolumn{5}{@{}p{120mm}}{Expressions for ``$a,\dots,
f$''
are Eqs.(36).}
\end{tabular*}
\end{table*}

\begin{table*}[h]
\setlength{\tabcolsep}{2.pc}
\newlength{\digitwidth} \settowidth{\digitwidth}{\rm 0}
\catcode`?=\active \def?{\kern\digitwidth}
\caption{}
\label{tab1}
\begin{tabular*}{\textwidth}{@{}lllll}
\hline
                 & \multicolumn{2}{l}{Additional $f_M
\pm f_E
$ }
                 & \multicolumn{2}{l}{Additional $T^+
\pm T_5^+
$ }
\\
\cline{2-3} \cline{4-5}
                 & \multicolumn{1}{l}{Plus $g_+$}
                 & \multicolumn{1}{l}{Plus $g_{-}$}
                 & \multicolumn{1}{l}{Plus
$\tilde{g}_+$}
                 & \multicolumn{1}{l}{Plus $\tilde{g}_{-
}$}         \\
\hline
{$\bf  \; \; \Gamma$'s}                    &

&

&
&
                  $
$
\\
$\zeta$                                          & $
1+g \frac{{\cal
RE}(g_L^{*}g_{+})}{\left| g_L\right| ^2} $         & $
1-h
\left| \frac{g_{-} }{g_L} \right| ^2$
& $ 1 $
&
 $ \xi $
\\
$\sigma$                                      & $ \zeta
$
&  $          1-j
\left|
\frac{g_{-}}{g_L}\right| ^2$
& $ 1     $
&
$   1 $
\\
$\xi$                                             & $  1
$
&  $1-k \left|
\frac{g_{-}}{g_L}\right| ^2                          $
&  $  1 $
&  $\frac{\left| g_L\right| ^2-\left| \frac{m\tilde g_{-
}}{2\Lambda
}\right| ^2}{\left| g_L\right| ^2+\left| \frac{m\tilde
g_{-
}}{2\Lambda
}\right| ^2}$\\
{$\bf  \; \; \Gamma_{LT}$'s }       &

&

&
&
                 $
$
\\
$\omega$                                     &  $
1-l \frac{{\cal
RE}(g_L^{*}g_{+})}{\left| g_L\right| ^2}$         & $
1-m
\left| \frac{g_{-}}{g_L}\right| ^2$
& $
1  $                    &
 $    \xi $
\\
$\eta$                                          &
$\omega
$
& $        1-n \left|
\frac{g_{-}}{g_L}\right| ^2$
& $  1 $
&
$   1    $
\\
$\omega'$                                   & $-
o \frac{{\cal
IM}(g_L^{*}g_{+})}{\left| g_L\right|^2}$           &  $
0
$
&    $ 0 $
&
$  0 $
\\
$\eta'$                                        & $
\omega'  $

&  $  0
$
&     $ 0 $
&
$   0 $
\\
\hline
\multicolumn{5}{@{}p{120mm}}{Expressions for ``$g,\dots,
o$''
are Eqs.(37).}
\end{tabular*}
\end{table*}

\begin{table*}[t]
\setlength{\tabcolsep}{1.5pc}
\newlength{\digitwidth} \settowidth{\digitwidth}{\rm 0}
\catcode`?=\active \def?{\kern\digitwidth}
\caption{}
\label{tabcpa}
\begin{tabular*}{\textwidth}{@{}l@{\extracolsep{\fill}}r
rr}
\hline
\\ [.5pt]
$E_{cm}$ & $\sigma (r_a)$                    &$\sigma
(\~{\beta})\simeq
\sigma
(\beta_a)$&\sigma (\beta^{\prime})$\\
                    &{\bf $CP\~{T_{FS}},CP$}&{\bf
$CP$\hspace{2pc}$\~{T_{FS}$
}        &{\bf $CP\~{T_{FS}},CP$}\\
\hline
$M_Z$       & $0.3\% $                             &
$\sim 10^{o}$
&  $\sim 15^{o}$ \\
$10 GeV$  & $0.05\% $                           & $\sim
3^{o}$
& $\sim  3^{o}$ \\
$4 GeV$    & $0.05\% $                           & $\sim
4^{o}$
& $\sim  5^{o}$ \\
\hline
\end{tabular*}
\end{table*}

\end{document}